\documentclass{svmult}

\usepackage{makeidx}         
\usepackage{graphicx}        
\usepackage{multicol}        
\usepackage[bottom]{footmisc}
\usepackage{subfigure}

\title*{An Experimental Study of Pedestrian Congestions: Influence of Bottleneck Width and Length}
\titlerunning{An Experimental Study of Pedestrian Congestions}
\authorrunning{J. Liddle et al.}
\author{Jack Liddle\inst{1}, Armin Seyfried\inst{1}, Wolfram Klingsch\inst{2}, Tobias Rupprecht\inst{2}, Andreas Schadschneider\inst{3} and Andreas Winkens\inst{2}}
\institute{
      J\"ulich Supercomputing Centre, Forschungszentrum J\"ulich. 
         \texttt{j.liddle@fz-juelich.de}, \texttt{a.seyfried@fz-juelich.de}
   \and
      Institute for Building Material Technology and Fire Safety, Universit\"at Wuppertal.
         \texttt{klingsch@uni-wuppertal.de}, \texttt{rupprecht@uni-wuppertal.de}, \texttt{winkens@uni-wuppertal.de}
   \and
   Institut f\"ur Theoretische Physik, Universit\"at zu K\"oln.
         \texttt{as@thp.uni-koeln.de}
               }
\begin{document}
\maketitle

\begin{abstract}
The placement and dimensioning of exit routes is informed by experimental data and theoretical models.  The experimental data is still to a large extent uncertain and contradictory.  In this contribution an attempt is made to understand and reconcile these differences with our own experiments.

\end{abstract}

\section{Introduction}

An understanding of pedestrian motion within high occupancy buildings is necessary in order to ensure safe and prompt evacuations.  Bottlenecks are an environment of interest and several factors influence this motion, namely the geometry of the bottleneck (width, length and location, the initial distribution of the pedestrians (densely/lightly packed, starting point) and sociological effects (group composition, motivation, stress levels).  In~\cite{Seyfried2009} several experiments examining pedestrian flow through bottlenecks were compared and contrasted with each other.  Between the experiments large discrepancies in the measurement of the flow were seen.  In~\cite{Hoogendoorn2005} the zippering effect was observed and the flow increased in a stepwise manner.

The size of the discrepancies casts doubts over the specification of planning regulations and guidelines, and as such we are compelled to reach an understanding of the underlying causes of these discrepancies.  This contribution attempts to address the underlying causes of these differences.

\begin{figure}
   \centering
   \subfigure[Kretz~\cite{Kretz2006a}.]
   {
      \includegraphics[scale=0.45]{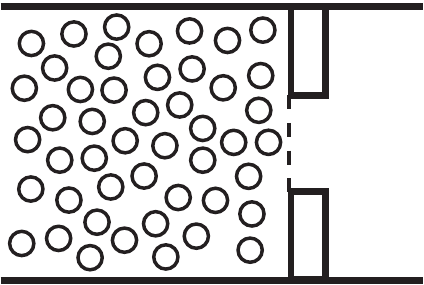}
      \label{fig:sketch_kretz}
   }
   \subfigure[Muir~\cite{Muir1996}.]
   {
      \includegraphics[scale=0.45]{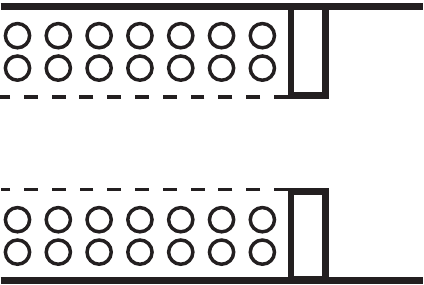}
      \label{fig:sketch_muir}
   }
   \subfigure[M\"uller~\cite{Mueller1981}.]
   {
      \includegraphics[scale=0.45]{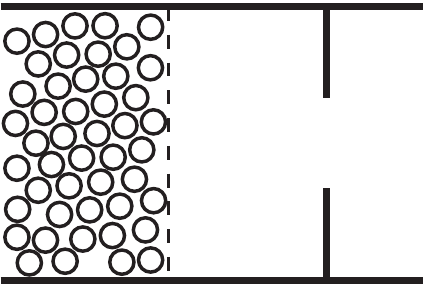}
      \label{fig:sketch_mueller}
   }
   \subfigure[Nagai~\cite{Nagai2006}.]
   {
      \includegraphics[scale=0.45]{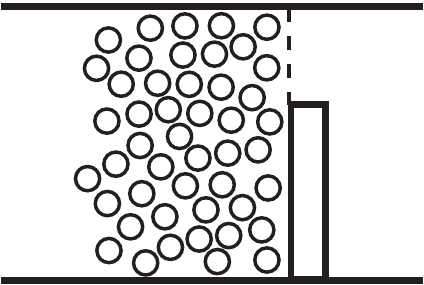}
      \label{fig:sketch_nagai}
   }
   \subfigure[Seyfried~\cite{Seyfried2009}.]
   {
      \includegraphics[scale=0.45]{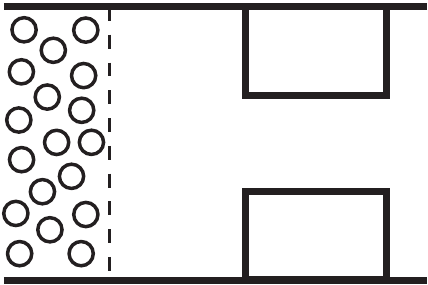}
      \label{fig:sketch_seyfried}
   }

   \caption{Comparison of experimental setups.}
   \label{fig:setups}
\end{figure}

In figure~\ref{fig:setups} sketches of the experimental setups discussed in~\cite{Seyfried2009} are shown.  The most immediately identifiable differences are the geometry of bottleneck and the initial distribution of the pedestrians.  We consider only the experiments from Kretz~\cite{Kretz2006a}, M\"uller~\cite{Mueller1981} and Seyfried~\cite{Seyfried2009} (figure~\ref{fig:sketch_kretz},~\ref{fig:sketch_mueller},~\ref{fig:sketch_seyfried}) in this analysis as they concern the motion of an evenly distributed group through a symmetrical bottleneck.

\begin{figure}
      \centering
      \subfigure[Still taken from experiment.]
      {
         \includegraphics[scale=0.35]{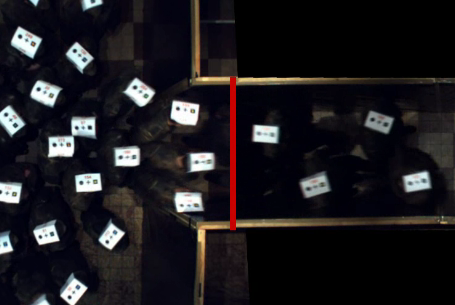}
         \label{fig:experimental-still}
      }
      \subfigure[Experimental setup.]
      {
         \includegraphics[scale=0.42]{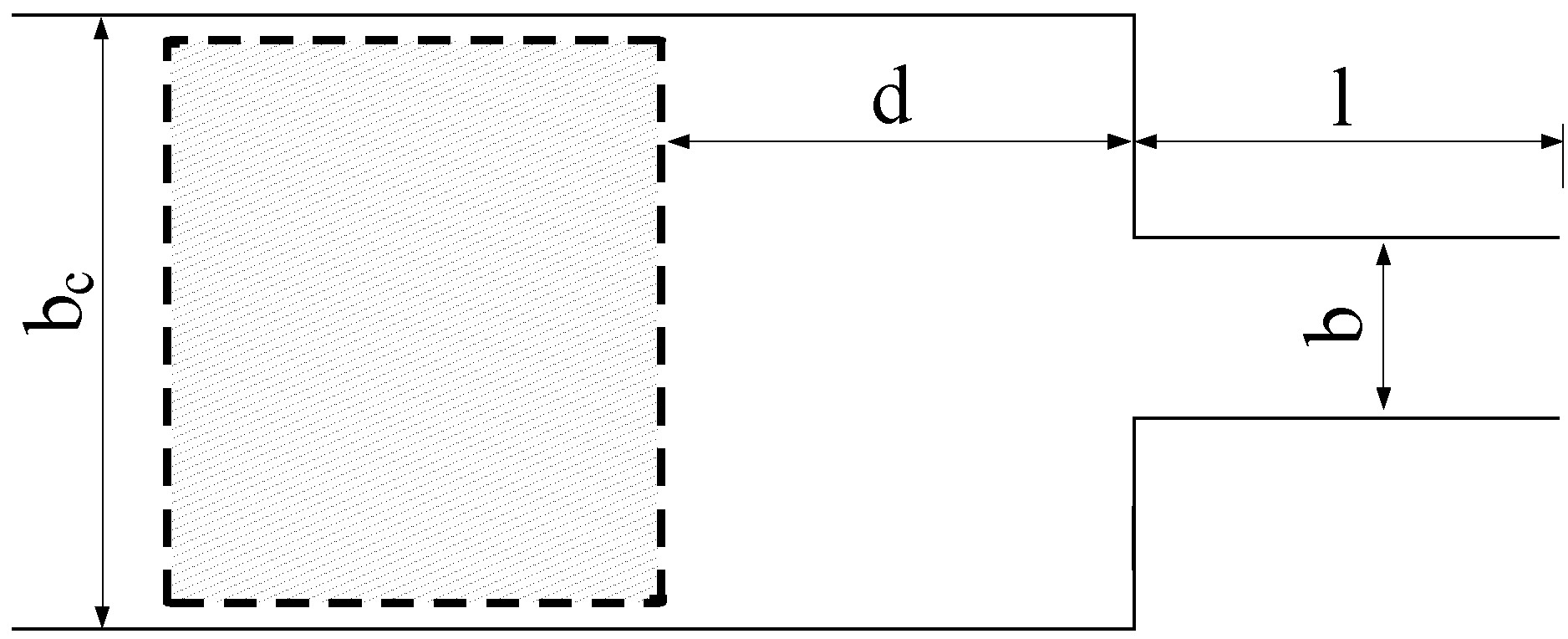}
         \label{fig:experimental-setup}
      }
      \caption{The experimental setup.}
   \end{figure}

Our experiment was performed in 2006 in the wardroom of the ‘Bergische Kaserne D\"usseldorf’.  The experimental setup allows the influence of the bottleneck width and length to be probed.  In one experiment the width, $b$, was varied (from $90$ to $250$ cm) at fixed corridor length, the other experiment investigated varying the corridor length, $l$, ($6,200,400$ cm) at fixed $b$ of $120$ cm.  Wider bottlenecks with more test persons were studied than in previous attempts.  The test group was comprised of soldiers.  Figure~\ref{fig:experimental-setup} shows a sketch of the experimental setup used to analyze the flow through bottlenecks, figure \ref{fig:experimental-still} shows a still taken from the experiment.  To ensure an equal initial density for every run, holding areas were marked on the floor (dashed regions). All together 99 runs with up to 250 people were performed over five days.

\section{Results}

\begin{figure}
   \centering
   \subfigure[Dependence on $b$.]
      {
         \includegraphics[scale=0.25]{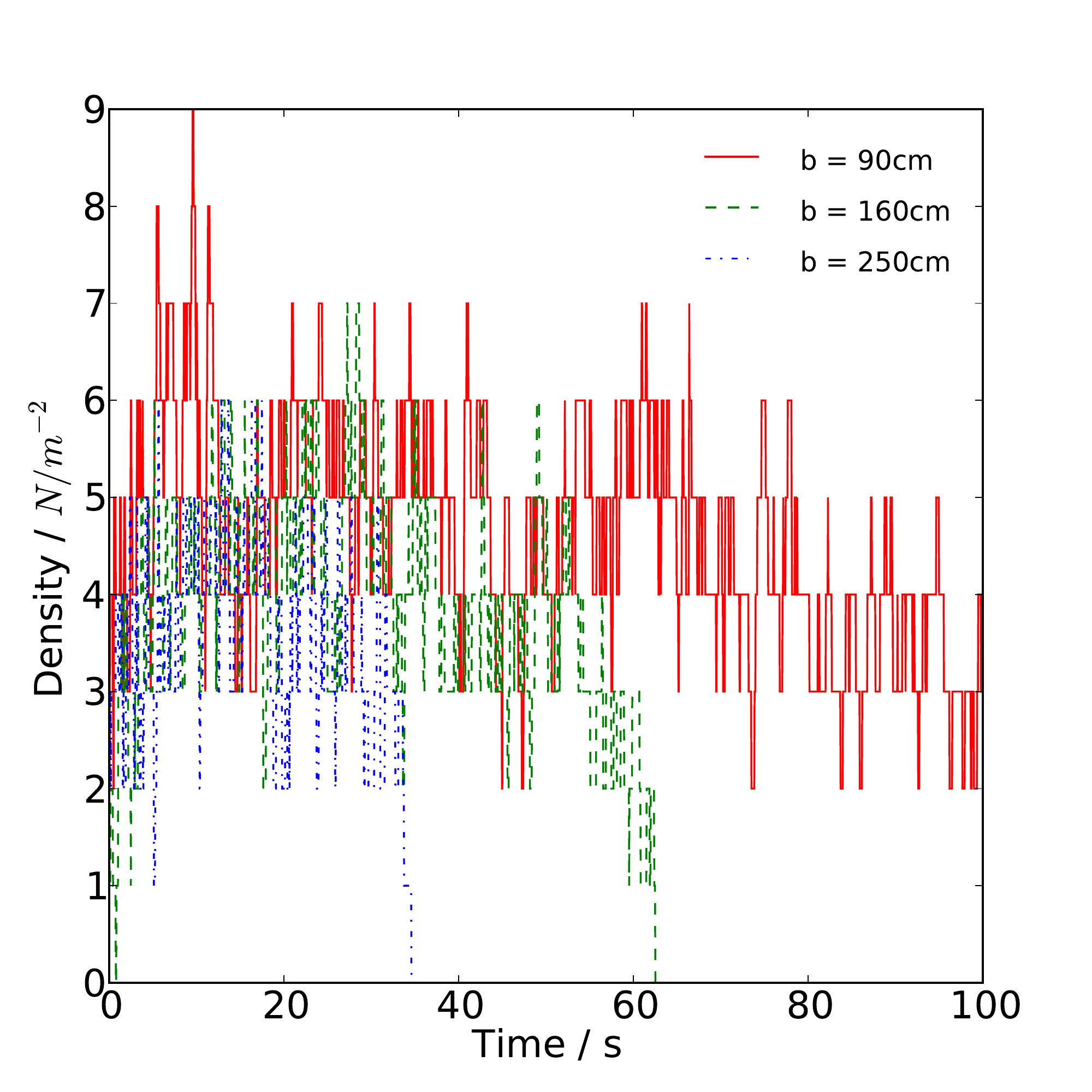}
         \label{fig:dens_time_b}
      }  
   \subfigure[Dependence on $l$.]
      {
         \includegraphics[scale=0.25]{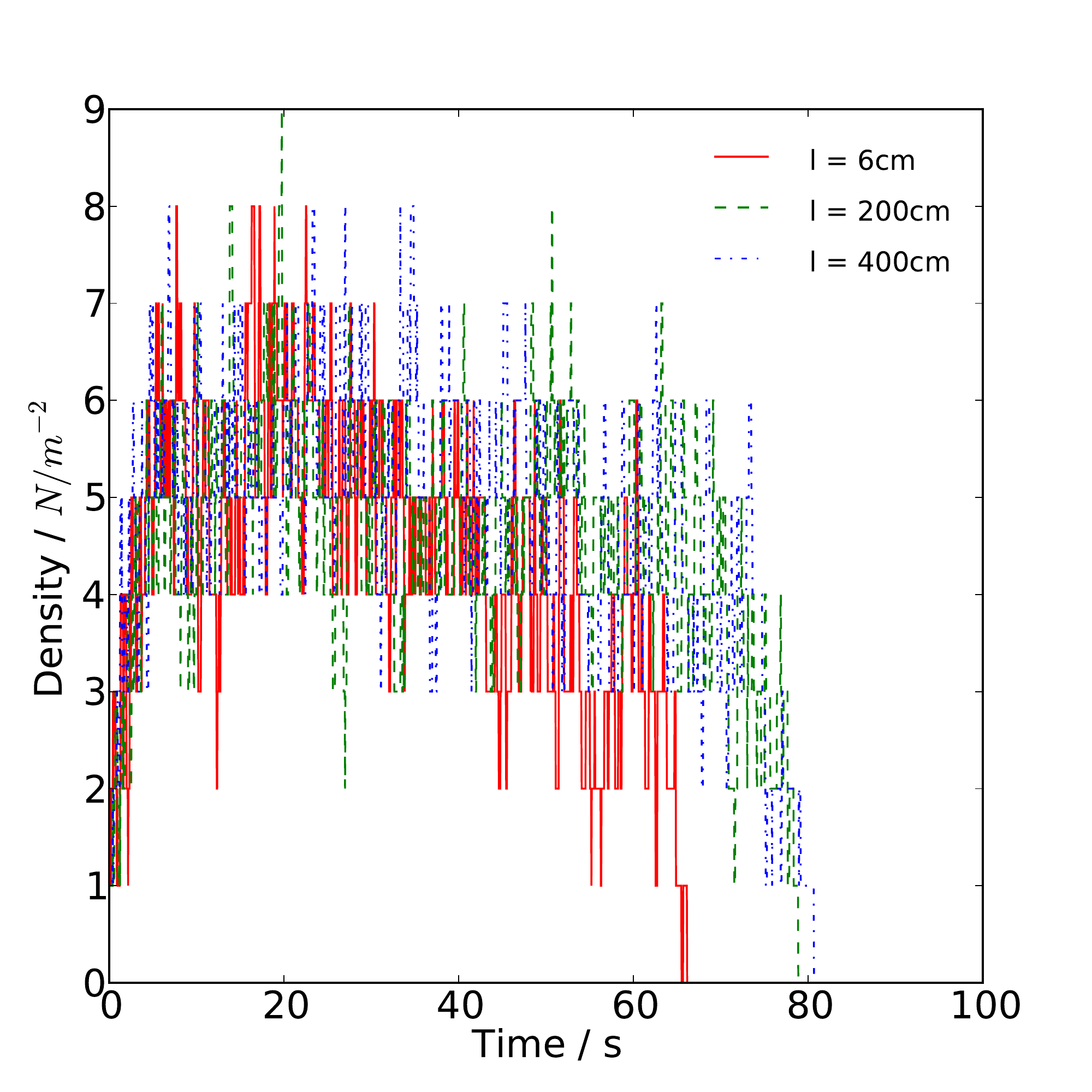}
         \label{fig:dens_time_l}
      }
  \caption{Time evolution of the density in front of bottleneck.}
  \label{fig:dens_time}
\end{figure}

The time evolution of the density in front of the bottleneck is shown in figure~\ref{fig:dens_time}.  In the bottleneck width experiment the dependence on $b$ is immediately obvious, figure~\ref{fig:dens_time_b}, with the highest densities being seen in front of the narrowest bottlenecks (between $20$ and $30$ seconds the densities are $5.46 \pm 0.75$, $4.79 \pm 1.00$ and $3.52 \pm 0.87$ for $b = 90,160,250$ cm respectively).  For the length experiments no appreciable difference in the densities can be observed.

\begin{figure}
   \begin{minipage}[b]{.5\linewidth}
   \subfigure[$b = 100$ cm, $l = 400$ cm.]
   {
      \includegraphics[scale=0.3]{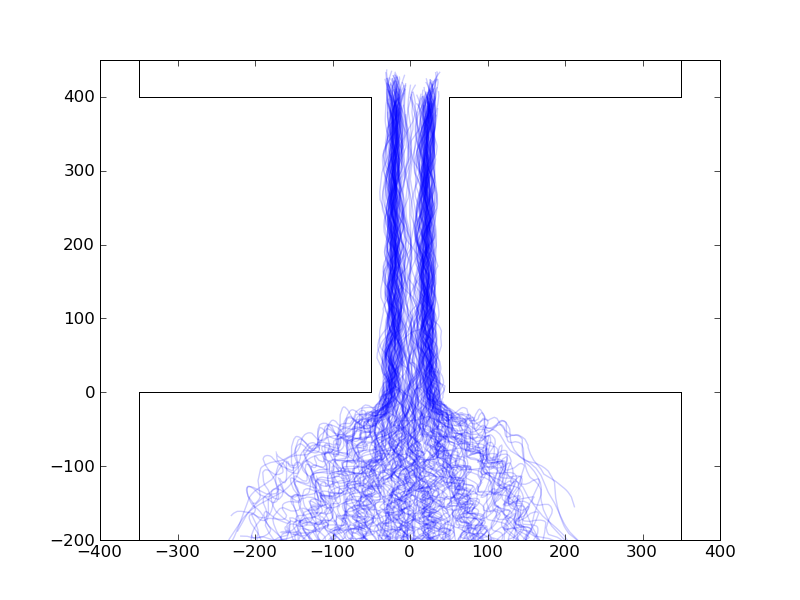}
      \label{fig:b100_paths}
   }
   \subfigure[$b = 160$ cm, $l = 400$ cm.]
   {
      \includegraphics[scale=0.3]{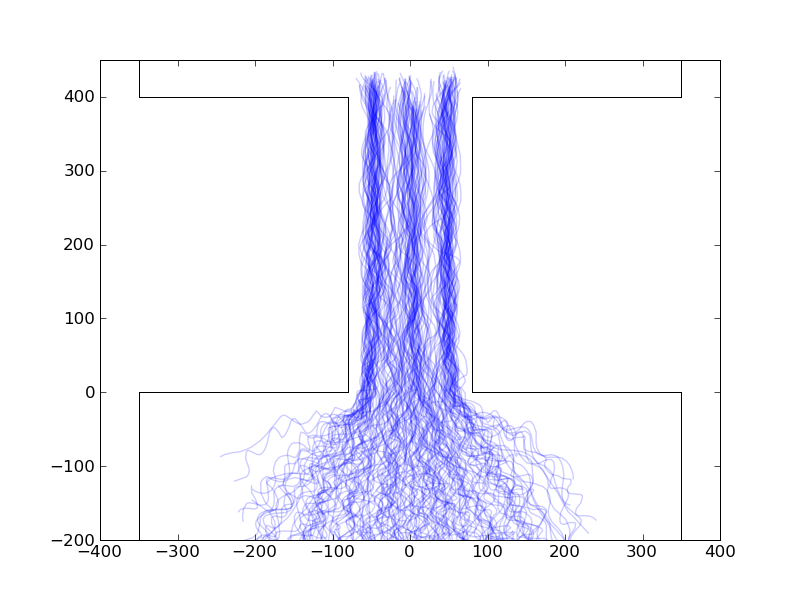}
      \label{fig:b160_paths}
   }
   \subfigure[$b = 200$ cm, $l = 400$ cm.]
   {
      \includegraphics[scale=0.3]{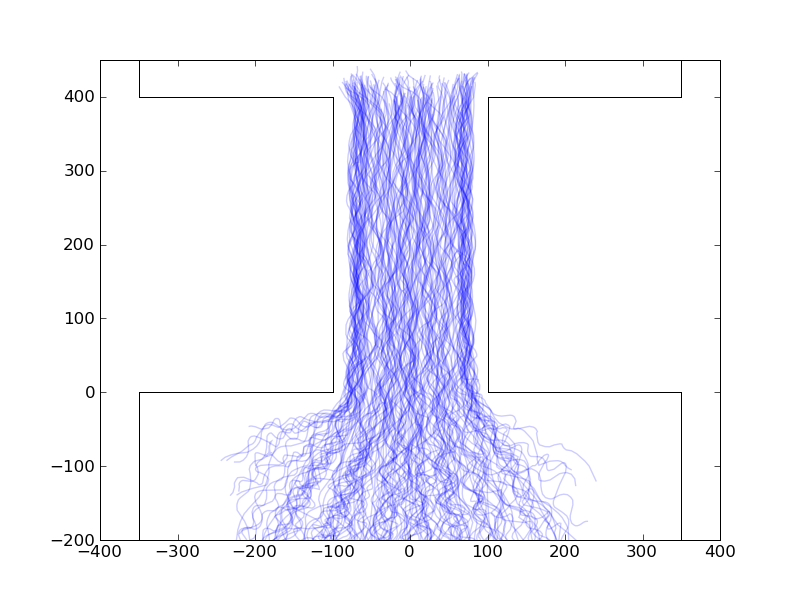}
      \label{fig:b200_paths}
   }
   \end{minipage}
   \begin{minipage}[b]{.5\linewidth}
   \subfigure[$b = 120$ cm, $l = 6$ cm.]
   {
      \includegraphics[scale=0.3]{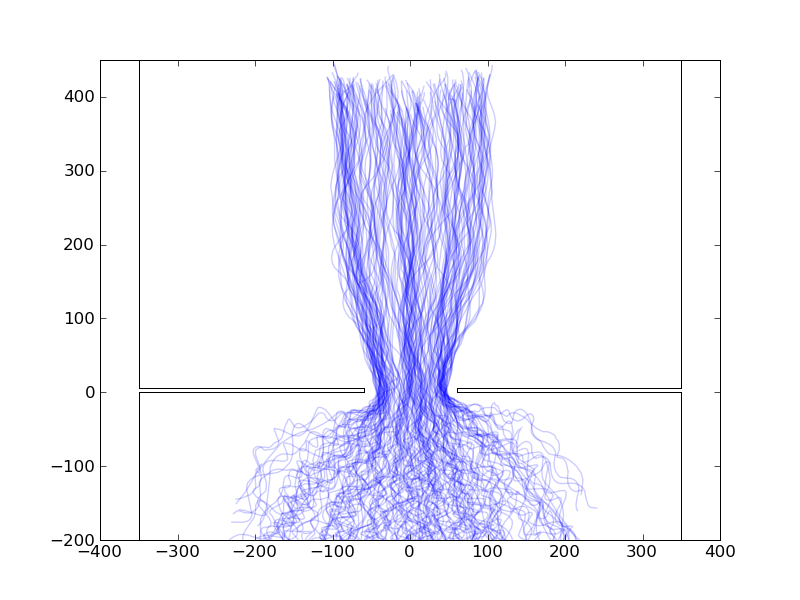}
   }
   \subfigure[$b = 120$ cm, $l = 200$ cm.]
   {
      \includegraphics[scale=0.3]{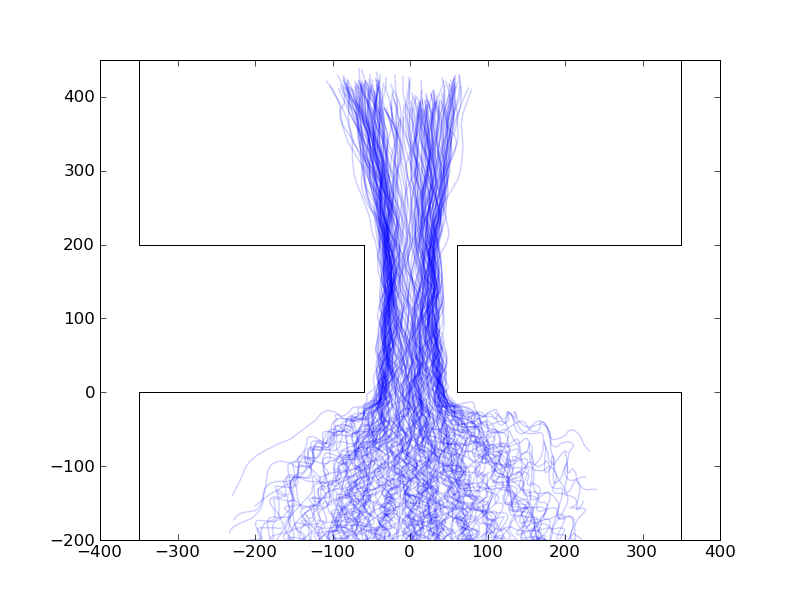}
   }
   \subfigure[$b = 120$ cm, $l = 400$ cm.]
   {
      \includegraphics[scale=0.3]{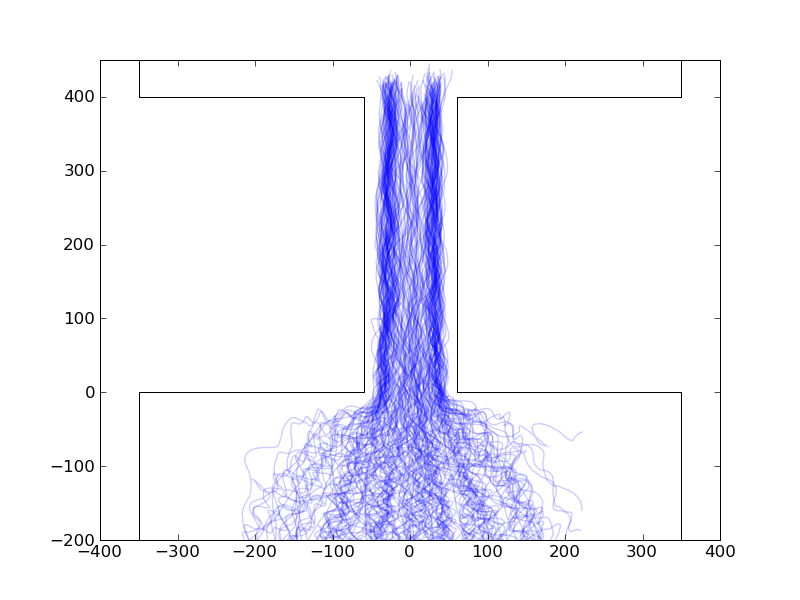}
   }
   \end{minipage}
   \caption{All pedestrian paths.}
   \label{fig:allpaths}
\end{figure}

With the accurate trajectories available to us, we can examine the process of lane formation in detail.  In figure~\ref{fig:allpaths} all the paths taken by pedestrians passing through bottlenecks of varying widths are overlayed.  From these various insights into the process of lane formation can be obtained.  The most immediate observation is that lanes are not formed continuously as the bottleneck width increases, figures~\ref{fig:b100_paths} to~\ref{fig:b200_paths}, instead lanes are formed along the walls of the bottleneck.  At fixed $b$, the path layouts for  $l = 200,400$ cm look qualitatively the same, with two lanes forming.  This contrasts with the $l = 6$ cm situation where three lanes are formed.

\begin{figure}
   \subfigure[$N$.]
   {
      \includegraphics[scale=0.28]{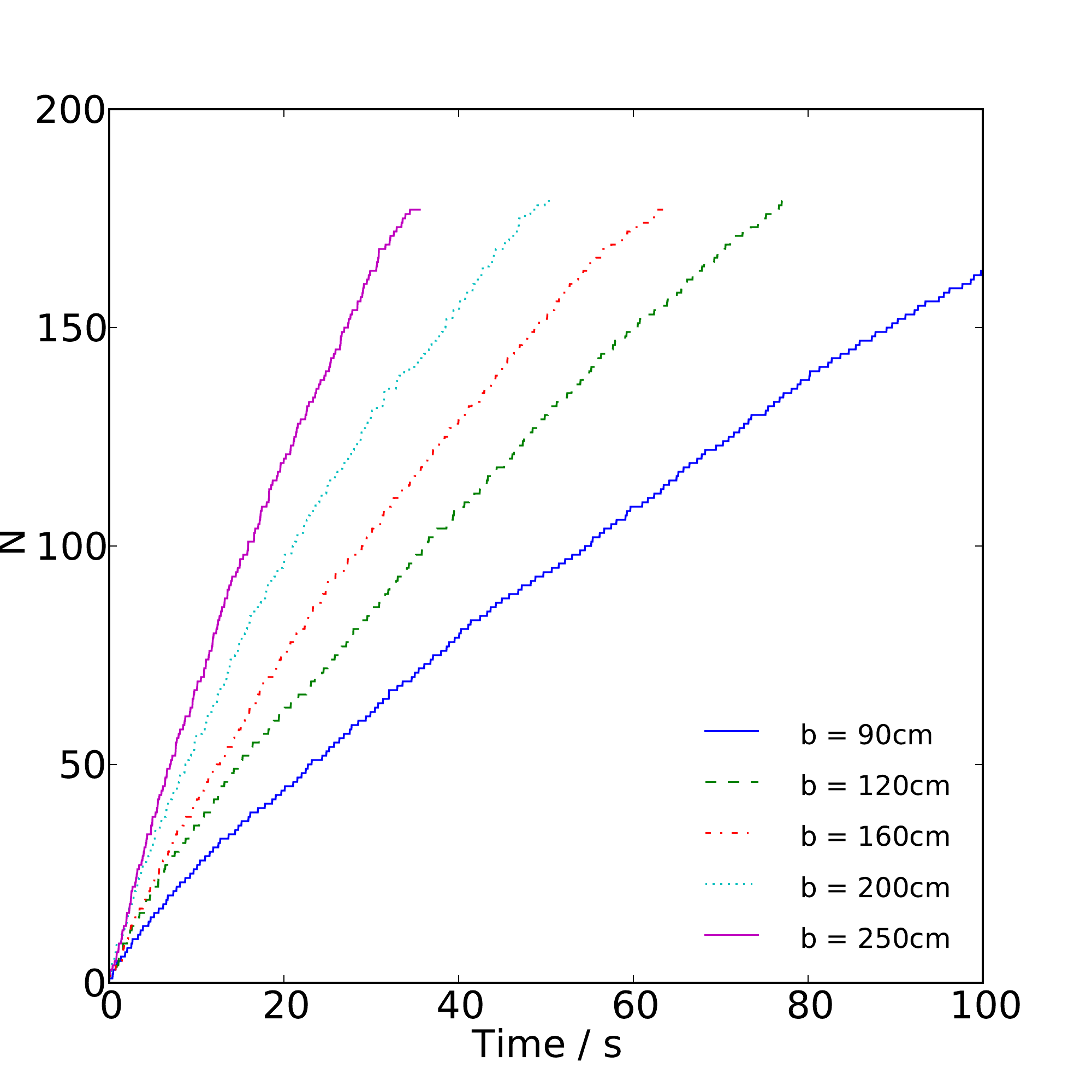}
      \label{fig:N_b}
   }
   \subfigure[Flow.]
   {
      \includegraphics[scale=0.28]{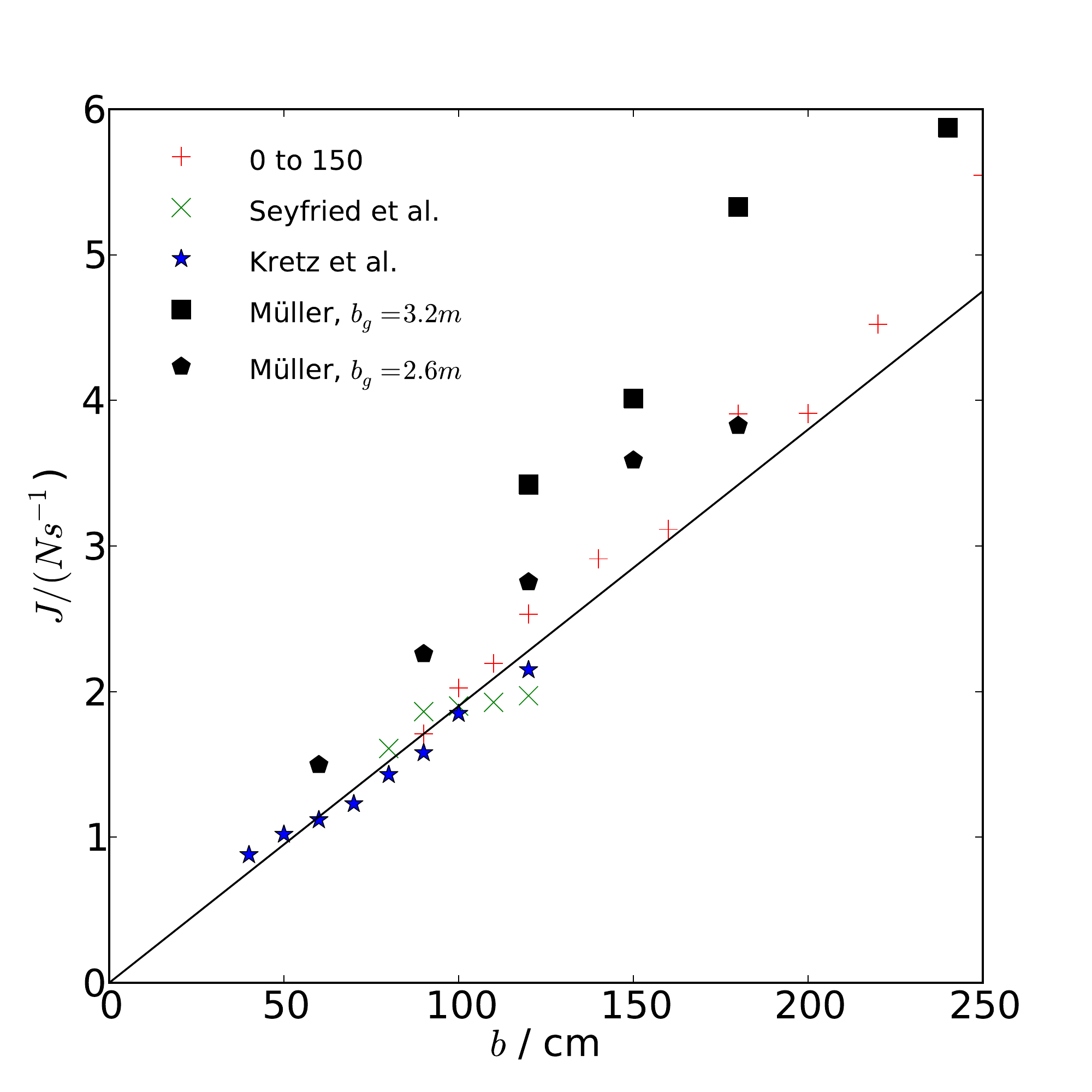}
      \label{fig:Flow_b}
   }
   \caption{$N$ the total number of pedestrians and the flow $J$, variation with $b$.}
   \label{fig:b_flows}
\end{figure}

\begin{figure}
   \subfigure[$N$.]
   {
      \includegraphics[scale=0.28]{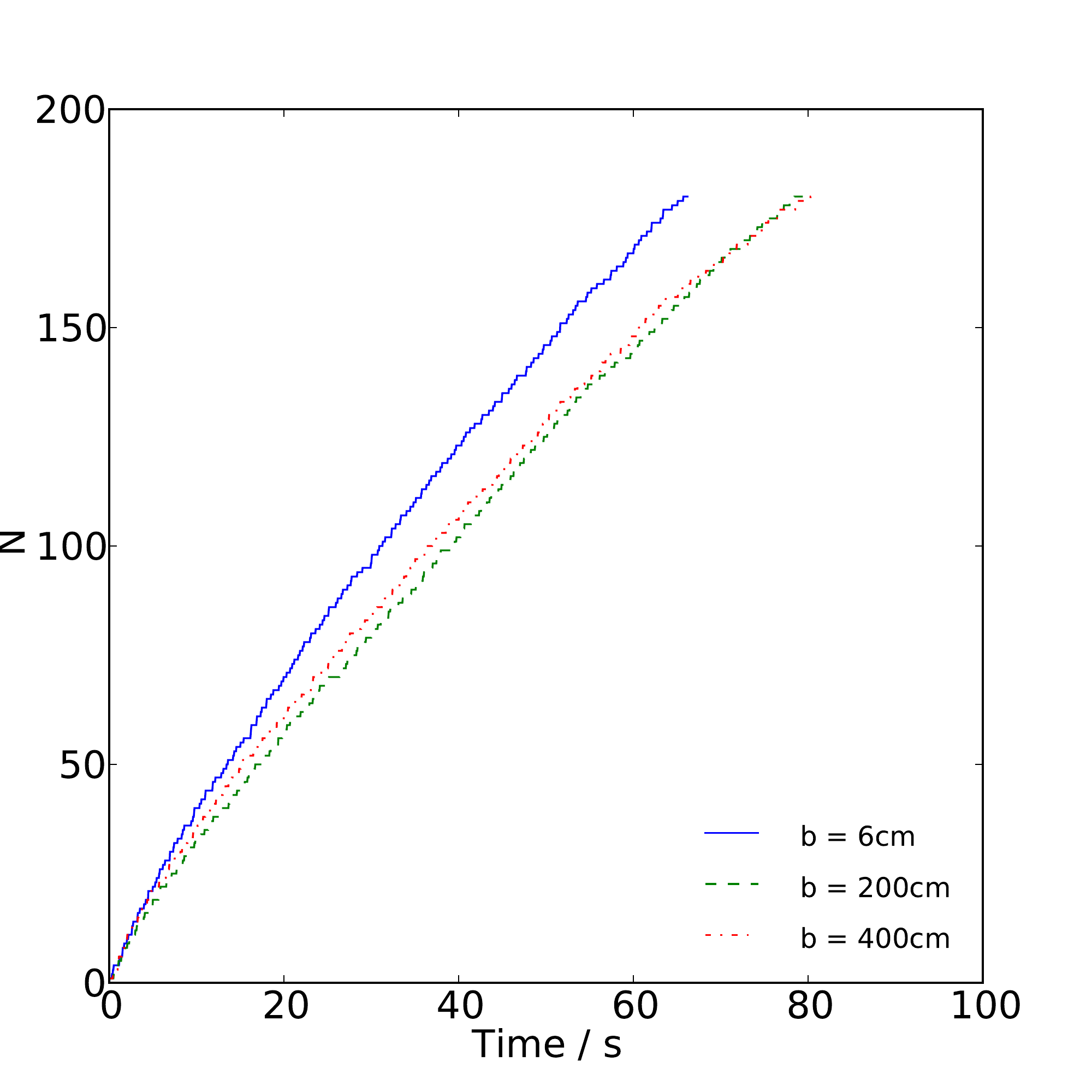}
      \label{fig:N_l}
   }
   \subfigure[Flow.]
   {
      \includegraphics[scale=0.28]{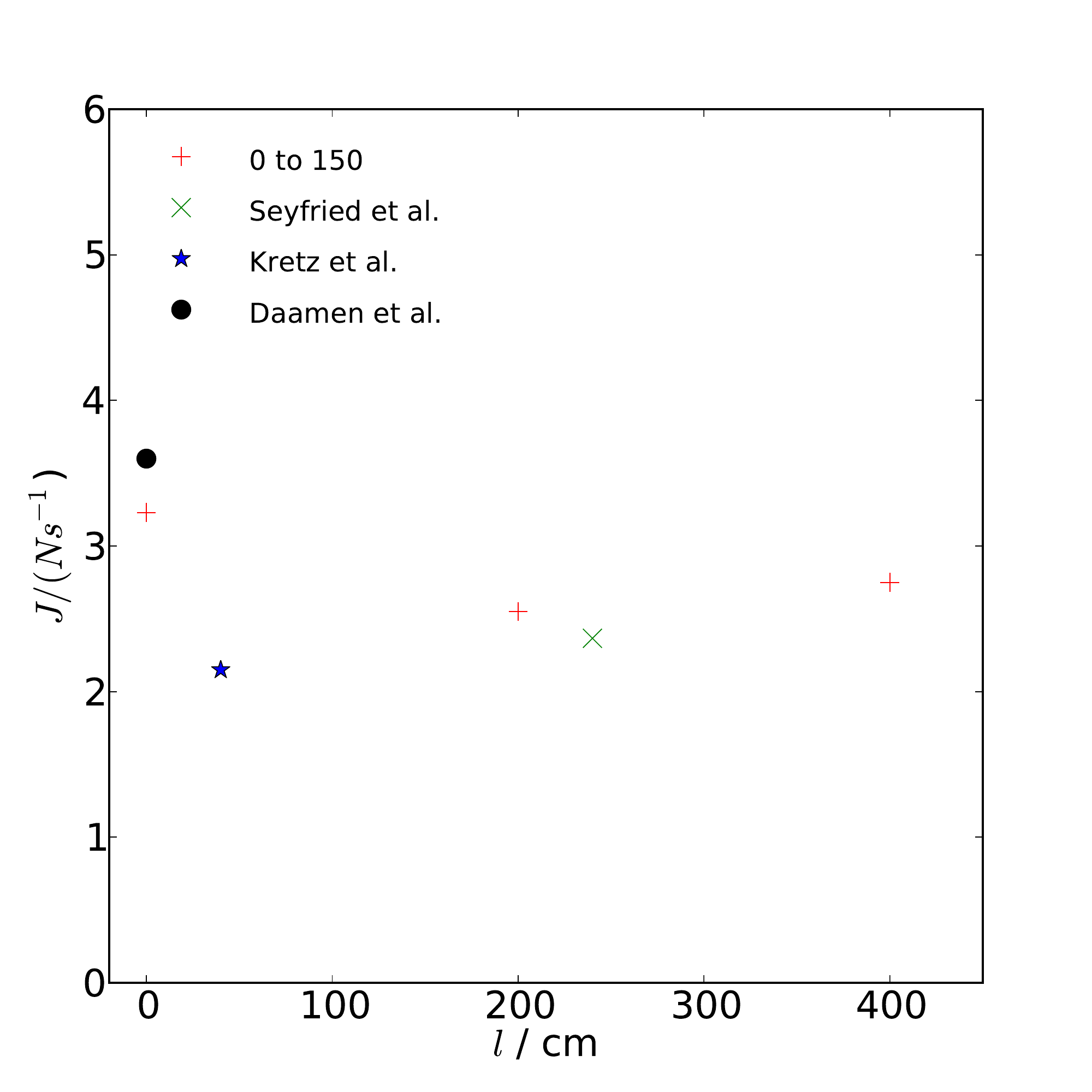}
      \label{fig:Flow_l}
   }
   \caption{$N$ the total number of pedestrians and the flow $J$, variation with $l$.}
   \label{fig:l_flows}

\end{figure}

In figure~\ref{fig:N_b} and~\ref{fig:N_l} $N$ can be seen to curve downwards therefore the flow is time-dependent, diminishing as the experiment runs its course.  The flow will depend on the number of pedestrians considered.  In this analysis we calculate the flow using the first 150 people.  Previous experiments have not observed this time dependence, as the participation in the experiments was not high enough (for example in~\cite{DaamenHoogendoorn-Delft} only 100 pedestrians took part).  As expected the flow exhibits a strong dependence on the width of the bottleneck $b$, see figure~\ref{fig:b_flows}.  The bottleneck length $l$ exerts virtually no influence on the flow, except for the case of the an extremely short constriction, figure~\ref{fig:b_flows}, where three lanes can be formed.  In figures~\ref{fig:Flow_b} and~\ref{fig:Flow_l}, the flow from our experiment is presented alongside previous measurements.  The black line in figure~\ref{fig:Flow_b} represents a constant specific flow of $1.9$ $m^{-1} s^{-1}$.  The difference between the flow at $l = 6$ cm and $l = 200,400$ cm is $\Delta J \simeq 0.5$ $s^{-1}$

The M\"uller datapoints lie significantly above the black line.  The M\"uller experimental setup features a large initial density of around $6$ $Pm^{-2}$ and an extremely short corridor.  The discrepancy between the M\"uller data and the empirical $J = 1.9 b$ line is roughly $\Delta J \simeq 0.5$ $s^{-1}$.  This difference can be accounted for due to the short corridor, but may also be due to the higher initial density in the M\"uller experiment.

\section{Stationary state}
It is reasonable to assume that a stationary state would be reached given greater participation, as the initial conditions are forgotten by the system.  However, from our data, we cannot detect any limiting value for the flow and therefore must presume that the participation level required for the system to reach a stationary state will be prohibitively high.  It is possible that shorter bottlenecks require longer to reach a stationary state and secondly eager people are more likely to position themselves at the front of the pack.  The individual eagerness could be controlled by deliberately shuffling the participants in the experiment.
\section{Conclusions and future prospects}

Pedestrian motion is influenced by innumerable factors.  Here we have shown that the length of the bottleneck exerts a considerable influence on the flow, and can account for some of the observed differences.  As expected the wider the bottleneck the greater the flow, also shorter bottlenecks permit a greater flow than longer ones.  We have also observed that inside the extremely short bottleneck three lanes are formed.  The influence of initial conditions on the time evolution of the system is still not understood.  No evidence of a stepwise increase in the flow was seen as predicted by the zippering effect.

Due to the large fluctuations inherent in the measurement of the density, it is difficult to recognise when a stationary state has been achieved.  Work is in progress on a measurement method which mitigates this problem.

\section{Acknowledgments}
The project is partially funded by the German Research Foundation (DFG) under the grant numbers KL 1873/1-1 and SE 1789/1-1.

\bibliographystyle{unsrt}
\bibliography{tgfpaper}
\end{document}